\documentclass[english,12pt,journal,draftclsnofoot,onecolumn]{IEEEtran}
\usepackage[T1]{fontenc}
\usepackage{amsmath}
\usepackage{graphicx}
\usepackage{setspace}
\usepackage{amssymb}

\makeatletter

\providecommand{\tabularnewline}{\\}

 \newtheorem{thm}{Theorem}
 \newtheorem{example}{Example}

\title{Estimation of Bit and Frame Error Rates of Low-Density Parity-Check Codes on Binary Symmetric Channels}

\usepackage{babel}
\makeatother
\begin{document}

\title{Estimation of Bit and Frame Error Rates of Low-Density Parity-Check
Codes on Binary Symmetric Channels}

\maketitle
\begin{abstract}
A method for estimating the performance of low-density parity-check
(LDPC) codes decoded by hard-decision iterative decoding algorithms
on binary symmetric channels (BSC) is proposed. Based on the enumeration
of the smallest weight error patterns that cannot be all corrected
by the decoder, this method estimates both the frame error rate (FER)
and the bit error rate (BER) of a given LDPC code with very good precision
for all crossover probabilities of practical interest. Through a number
of examples, we show that the proposed method can be effectively applied
to both regular and irregular LDPC codes and to a variety of hard-decision
iterative decoding algorithms. Compared with the conventional Monte
Carlo simulation, the proposed method has a much smaller computational
complexity, particularly for lower error rates.
\end{abstract}

\section{Introduction}

\PARstart{L}{ow}-density parity-check (LDPC) codes, which were first
proposed by Gallager \cite{key-12}, have attracted considerable attention
over the past decade due to their capacity-achieving error performance
and low complexity of the associated iterative decoding algorithms.
Iterative decoding algorithms are particularly simple if binary messages
are used. Such algorithms, which are referred to as hard-decision
iterative algorithms, are the subject of this paper. Examples are
the so-called Gallager algorithms A (GA) and B (GB) \cite{key-12,key-3,key-5},
their variants \cite{key-6} and majority-based (MB) algorithms \cite{key-7}.

The asymptotic performance analysis of LDPC codes under iterative
decoding algorithms is now a rather mature subject. On the other hand,
the finite-length analysis of LDPC codes is still an active area of
research with many open problems. In this paper, our focus is on practical
LDPC codes with finite block lengths in the range of a few hundreds
to a few thousands of bits.

In \cite{key-8}, Di \emph{et al.} studied the average performance
of the ensembles of regular finite-length LDPC codes over binary erasure
channels (BEC), by relating the decoding failures of the belief propagation
algorithm over a BEC to graphical objects called \emph{stopping sets}.
Another step in analyzing the performance of finite-length LDPC codes
was taken by Richardson \cite{key-9}, where the performance of a
\emph{given} LDPC code, particularly in the error floor region, was
related to the \emph{trapping sets} of the underlying graph. Most
recently, Cole \emph{et al.} \cite{key-13}, devised an efficient
algorithm for estimating the FER floors of soft-decision LDPC decoders
on the AWGN channel using importance sampling. Estimations of the
FER of GA and GB algorithms for certain categories of LDPC codes were
also presented in \cite{key-14}, \cite{key-15}. In an earlier paper
\cite{key-10}, we derived upper and lower bounds on the FER of LDPC
codes under hard-decision algorithms over a BSC. The bounds which
were tight for sufficiently small crossover probabilities of the channel
provided good estimates for the FER in this region. 

In this paper, we propose a method to estimate the performance of
LDPC codes decoded by hard-decision iterative algorithms over a BSC.
Compared to the methods of \cite{key-9,key-13,key-14,key-15} which
first identify the most relevant trapping sets and then evaluate their
contribution to the error floor, our approach does not require the
identification of the trapping sets and is thus much simpler. Moreover,
our approach is not only applicable to the error floor region, but
also provides very accurate estimates in the waterfall region. In
addition, we provide estimates for the BER, which is not even discussed
in \cite{key-9,key-13,key-14,key-15}. In fact one of the important
contributions of this work is to shed some light on the generally
complex structure of the error events for iterative decoding, and
to be able to establish simple approximate relationships between these
complex events and the smallest initial error patterns that the decoder
fails to correct. These relationships are the key in deriving the
estimates. 

The rest of the paper is organized as follows. In section II, we discuss
different decoder failures for hard-decision algorithms. We then propose
our FER and BER estimation methods and discuss their complexity in
Section III. Simulation results are given in Section IV. Section V
concludes the paper.

\section{Failures of Hard-Decision Iterative Decoding Algorithms}

Hard-decision iterative decoding algorithms are initiated by the outputs
from the BSC. Hard messages are then passed between the variable nodes
and the check nodes in the Tanner graph of the code through the edges
of the graph in both directions and iteratively. We assume that the
alphabet space for messages is $\{+1,-1\}$ and that the decoder is
symmetric \cite{key-3}. To evaluate the performance, we can then
assume that the all-one codeword is transmitted.

An important category of hard-decision iterative algorithms are majority-based
($\textrm{MB}{}^{\omega}$) algorithms, where $\omega$ is referred
to as the order of the algorithm. For an $\textrm{MB}{}^{\omega}$
algorithm, in iteration zero, check nodes are inactive and variable
nodes pass their channel messages. In iteration $l\geq1$, check nodes
pass the product of their extrinsic incoming messages in iteration
$l-1$, and variable nodes pass their channel messages unless at least
one-half plus $\omega$ of their extrinsic incoming messages in iteration
$l$ disagree with the corresponding channel messages, in which case
the negative of those (channel) messages are passed. In other words,
for a variable node with degree $d_{v}$ and channel message $m_{0}$,
$-m_{0}$ is passed if at least $\lceil d_{v}/2\rceil+\omega$ of
extrinsic incoming messages are $-m_{0}$, where $\left\lceil x\right\rceil $
is the smallest integer which is greater than or equal to $x$. The
order $\omega$ can be any non-negative integer less than or equal
to $d_{v}-1-\lceil d_{v}/2\rceil$. This maximum value corresponds
to GA. 

Decoder failures of iterative algorithms can be related to graphical
objects, called trapping sets \cite{key-9}. A trapping set is defined
as a set of variable nodes that cannot be eventually corrected by
the decoder. In general, trapping sets depend not only on the structure
of the graph but also on the decoder inputs (channel) and the decoding
algorithm. For example, while for the belief propagation algorithm
over the BEC, the trapping sets are precisely the stopping sets, for
the maximum likelihood decoding over the BEC or any other channel,
the trapping sets are the nonzero codewords. The structure of the
trapping sets of hard-decision iterative decoding algorithms on a
BSC is in general unknown. The following theorem identifies the structure
of certain trapping sets for MB algorithms. This theorem is in fact
a generalization of Facts 3 and 4 of \cite{key-9}. Facts 3 and 4
of \cite{key-9} only apply to GA and GB on regular Tanner graphs
with variable node degree 3 and check node degree 6. The following
theorem extends this to irregular codes with arbitrary degree distributions
where different nodes are allowed to perform different majority-based
algorithms. 

\begin{thm}
\label{thm1:Trapping Set Structure}For an LDPC code with degree distribution
pair ($\lambda$, $\rho$), suppose that variable node $j$ performs
MB of order $\omega_{j}$ denoted by MB$^{\omega_{j}}$ ($0\leq\omega_{j}\leq d_{j}-1-\lceil d_{j}/2\rceil$),
where $d_{j}$ is the degree of variable node $j$. Consider a set
of variable nodes $S$ and denote its induced subgraph in the code's
Tanner graph by $G(S)$. Also denote the check nodes which have an
odd degree in $G(S)$ by $C_{o}(S)$. If each variable node$\omega$
$j$ in the Tanner graph has at most $\lceil d_{j}/2\rceil+\omega_{j}-1$
neighbors from $C_{o}(S)$, then $S$ is a trapping set. 
\end{thm}
\begin{proof}
We show that if all the variable nodes in $S$ are initially in error
and all the other nodes are initially correct, then the decoder fails
to correct the errors and the initial error pattern remains unchanged
throughout the iterations. 

In the first iteration, the outgoing messages of a variable node $j$
is the initial message $m_{0}\in\{-1,1\}$ of node $j$. Based on
the updating rule for MB algorithms in the check nodes, it is easy
to see that node $j$ receives $-m_{0}$ along all the edges connected
to $C_{o}(S)$ and $m_{0}$ along all the edges connected to $\overline{C_{o}(S)}=C\setminus C_{o}(S)$,
where $C$ is the set of all the check nodes in the Tanner graph.
This is regardless of whether node $j$ is in the set $S$ or not.
Due to the structure of $S$, node $j$ will therefore have at most
$\lceil d_{j}/2\rceil+\omega_{j}-1$ incoming messages with their
value equal to $-m_{0}$. This implies that for each outgoing message
of node $j$ at the start of the second iteration, the number of extrinsic
incoming messages with value $-m_{0}$ is also at most $\lceil d_{j}/2\rceil+\omega_{j}-1$.
Hence the value of the outgoing message will remain $m_{0}$. This
is because, for the outgoing message to change, MB$^{\omega_{j}}$
requires at least $\lceil d_{j}/2\rceil+\omega_{j}$ extrinsic incoming
messages with value $-m_{0}$. This means that all the messages exchanged
between variable nodes and check nodes, remain unchanged throughout
the iterations. This implies that the decoded values of the variable
nodes remain the same as the initial messages regardless of the number
of iterations.
\end{proof}
Theorem \ref{thm1:Trapping Set Structure} describes an instance of
one type of trapping set which we refer to as \emph{fixed-pattern}.
For hard-decision algorithms in general, however, this is not the
only type of trapping sets. In our experiments with different LDPC
codes and different hard-decision iterative decoding algorithms, we
have observed the following types of decoder failures corresponding
to different types of trapping sets. The observations are made by
tracking the error positions at the output of the decoder throughout
iterations.

\begin{enumerate}
\item Fixed-pattern: After a finite number of iterations, the error positions
at the output of the decoder remain unchanged.
\item Oscillatory-pattern: After a finite number of iterations, the error
positions at the output of the decoder oscillate periodically within
a small set of variable nodes. 
\item Random-like: Error positions change with iterations in a seemingly
random fashion. The errors seem to propagate in the Tanner graph and
result in a larger number of errors at the output of the decoder even
if the initial error pattern has only a small weight.
\end{enumerate}
Note that in the first two cases, there could be a transition phase,
through which the error pattern settles into the steady-state. Also
noteworthy is that in all the random-like cases that we have observed,
the trapping set includes all the bits in the codeword, i.e., no bit
can be eventually corrected by the decoder. In our experiments, we
also observe that the percentage of random-like failures out of the
total number of failures increases on average as the weight of the
initial error patterns increases. 

The relationship between the trapping sets and the structure of the
Tanner graph is in general complex. With the exception of Theorem
\ref{thm1:Trapping Set Structure} and code-specific results such
as those of \cite{key-14,key-15}, we are not aware of any other such
result. Our experiments show that in many cases cycles in the Tanner
graph are part of the trapping sets. The relationships among the cycles,
the trapping sets and the initial error patterns that trap the decoder
in the trapping sets, however, seem to be complex. For example, in
a Tanner graph with girth 6, while variable nodes in cycles of length
6 can form trapping sets for GA, they may be corrected by other MB
algorithms. Moreover, there are cases where the variable nodes in
a cycle of length 6 form a fixed-pattern trapping set for GA if these
nodes are initially in error with all the other nodes received correctly.
The same variable nodes, however, can be corrected by GA if the initial
error pattern also contains other variable nodes.

Due to the complexity of identifying and enumerating the different
types of trapping sets for a given LDPC code under a given hard-decision
iterative decoding algorithm, we take an approach different than that
of \cite{key-9,key-13,key-14,key-15}. Our approach is universal in
that it can be applied to any LDPC code with arbitrary degree distributions
and to any decoding algorithm as long as the complexity of implementation
is manageable. The algorithm is simply based on enumerating the initial
error patterns of smallest weight that cannot be all corrected by
the decoder. By using this information, we then estimate the contribution
of all the other initial error patterns with larger weights to the
total FER and BER.

\section{Error Rate Estimation}

\subsection{Frame Error Rate Estimation}

Consider a given LDPC code with block length $n$ decoded by a given
hard-decision iterative algorithm over a BSC with crossover probability
$\varepsilon$. Denote the set of all the error patterns of weight
$i$ by $S_{i}$, and those that cannot be corrected by the decoder
by $E_{i}$. Clearly, $|S_{i}|=\binom{n}{i}$. Suppose that the decoder
can correct all the error patterns of weight $J-1$ and smaller, i.e.,
$|E_{i}|=0,\:\forall i<J$. Also suppose that there are $|E_{J}|\neq0$
weight-$J$ error patterns that the decoder fails to correct. The
FER is then equal to \begin{eqnarray}
FER & = & {\displaystyle \sum_{i=J}^{n}P(e|i)p_{i}}={\displaystyle \sum_{i=J}^{n}\frac{|E_{i}|}{|S_{i}|}p_{i}}={\displaystyle \sum_{i=J}^{n}|E_{i}|\varepsilon^{i}(1-\varepsilon)^{n-i}},\label{eq:3}\end{eqnarray}
 where $e$ is the event of having a frame (codeword) decoded erroneously,
$i$ denotes the weight of the initial error pattern at the input
of the decoder, and $p_{i}$ is the probability of having $i$ errors
at the output of the channel (or the input of the decoder) given by
the binomial distribution $p_{i}=\binom{n}{i}\varepsilon^{i}(1-\varepsilon)^{n-i}$.

The first term of the summation in (\ref{eq:3}) is denoted by $P(J)$
and is equal to $|E_{J}|\varepsilon^{J}(1-\varepsilon)^{n-J}$. To
estimate the FER, we enumerate $E_{J}$ and calculate $P(J)$ precisely.
For the other terms in (\ref{eq:3}), we estimate $|E_{i}|$ as a
function of $|E_{J}|$ as follows.

For a given $i>J$, we partition the set $S_{i}$ as $S_{i}=S_{i}^{'}\cup\overline{S_{i}^{'}}$,
where $S_{i}^{'}$ is the set of error patterns of weight $i$, each
containing at least one element of $E_{J}$ as its sub-pattern, and
$\overline{S_{i}^{'}}$ is the complement set of $S_{i}^{'}$ in $S_{i}$.
The set $\overline{S_{i}^{'}}$ thus contains the elements of $S_{i}$
that do not have any elements of $E_{J}$ as their sub-patterns. We
make the following assumptions:

\emph{Assumption (a): No error pattern in the set $S_{i}^{'}$ can
be corrected by the decoder. }

\emph{Assumption (b):} \emph{Every error pattern in the set} $\overline{S_{i}^{'}}$
\emph{can be corrected by the decoder.}

By the above assumptions, we have $|E_{i}\cap S_{i}^{'}|=|S_{i}^{'}|$
and $|E_{i}\cap\overline{S_{i}^{'}}|=0$, and therefore $|E_{i}|=|E_{i}\cap S_{i}^{'}|+|E_{i}\cap\overline{S_{i}^{'}}|=|S_{i}^{'}|$.
The key in estimating $|E_{i}|,\: i>J$, is that $|S_{i}^{'}|$ can
be approximated as a function of $|E_{J}|$ using the following combinatorial
arguments. Consider the probability ${\cal {P}}$ that a randomly
selected weight-$i$ error pattern contains at least one element of
$E_{J}$ as its sub-pattern. We then have\begin{equation}
{\cal {P}}=\frac{|S_{i}^{'}|}{|S_{i}|}\approx1-\left(1-\frac{|E_{J}|}{|S_{J}|}\right)^{\binom{i}{J}}.\label{eq:4}\end{equation}
The first equality follows from the fact that there are $|S_{i}|$
equally likely possibilities out of which $|S_{i}^{'}|$ are the favorable
cases. The second part of the equation is a consequence of approximating
the random experiment by a sequence of $\binom{i}{J}$ independent
and identically distributed Bernoulli trials, each involving the selection
of a weight-$J$ error pattern with {}``success'' defined as the
pattern being in $E_{J}$. The probability of success is then $|E_{J}|/|S_{J}|$,
and the probability of having at least one success in $\binom{i}{J}$
trials is $1-(1-|E_{J}|/|S_{J}|)^{\binom{i}{J}}$ following the binomial
distribution. From (\ref{eq:4}), we have\begin{eqnarray}
|S_{i}^{'}| & \approx & |S_{i}|\left[1-\left(1-\frac{|E_{J}|}{|S_{J}|}\right)^{\binom{i}{J}}\right]\approx|S_{i}|\frac{|E_{J}|\binom{i}{J}}{|S_{J}|}=|E_{J}|\binom{n-J}{i-J},\label{eq:5}\end{eqnarray}
where the second approximation is obtained by considering only the
first two terms in the binomial expansion.

Note that Assumption (a) is valid for the belief propagation decoder
on a BEC. For hard-decision iterative decoding algorithms on a BSC,
although Assumptions (a) and (b) are not necessarily correct, they
are approximately valid, especially for smaller values of $i$. In
particular, our studies show that these assumptions are statistically
viable and result in good approximations for $|E_{i}|,\: i>J$.

Combining (\ref{eq:3}) and (\ref{eq:5}) with $|E_{i}|\approx|S_{i}^{'}|,\: i>J$,
we derive the following estimates for the FER: {}``Lower'' estimate,
\begin{equation}
FER_{L}(N)=P(J)+{\displaystyle \sum_{i=J+1}^{N}|E_{J}|\binom{n-J}{i-J}}\varepsilon^{i}(1-\varepsilon)^{n-i},\label{eq:FER_L}\end{equation}
{}``Upper'' estimate,\begin{eqnarray}
FER_{U}(N) & = & P(J)+{\displaystyle \sum_{i=J+1}^{N}|E_{J}|\binom{n-J}{i-J}}\varepsilon^{i}(1-\varepsilon)^{n-i}+\left[1-\sum_{i=0}^{N}\binom{n}{i}\varepsilon^{i}(1-\epsilon)^{n-i}\right],\label{eq:FER_U}\end{eqnarray}
where $N\in\{ J+1,J+2,\ldots,n\}$ is a parameter to be selected for
the best accuracy of the estimates. It is clear that $FER_{L}(N)\leq FER_{U}(N)$,
and the equality holds if and only if $N=n$. The difference between
the two estimates is the probability that the input error patterns
to the decoder have a weight larger than $N$. While $FER_{L}(N)$
is derived based on the assumption that error patterns of weight larger
than $N$ occur with negligible probability, $FER_{U}(N)$ is obtained
based on the assumption that for such input error patterns the decoder
fails with probability one. Our observations show that in practice
there exists a certain threshold $N_{0}$ for error weights, around
which a relatively abrupt change in the percentage of failures occurs.
This is such that for error weights larger than $N_{0}$, the probability
of failure goes to one very rapidly. 

Later in Section IV, we will see that $FER_{U}(N_{0})$ provides a
very accurate estimate of the FER for all channel crossover probabilities
of interest. This suggests the following practical approach for determining
$N_{0}$. We perform Monte Carlo simulations at high FER values, say
around $0.01-0.1$. We then choose $N_{0}$ such that the estimate
$FER_{U}(N_{0})$ is the closest to the simulated FER. As the Monte
Carlo simulations are performed at high FER values, their complexity
is low and easily manageable.

\subsection{Bit Error Rate Estimation}

At the first glance, the problem of BER estimation may seem very complicated
as we observe that even for the initial error patterns of the same
weight, depending on the type of the trapping set, the number of bits
in error at the end of the decoding can be quite different. Our further
study into this however reveals that despite this variety of possibilities,
one can estimate the average number of bit errors depending on the
initial weight $i$ of the error patterns. In general, we expect the
conditional average BER given the initial error weight $i$ to be
an increasing function of $i$. To simplify the analysis however,
we partition the range of $J\leq i\leq n$ into two subsets, $J\leq i<N_{0}$
and $N_{0}\leq i\leq n$. For these partitions, we estimate the average
number of bit errors by $J$ and $M$, respectively, where $M$ is
the estimate of the average number of bit errors for error patterns
of weight $N_{0}$, obtained by Monte Carlo simulations. We thus derive
our estimate for the BER based on the FER estimate $FER_{U}(N_{0})$,
proposed in the previous subsection, as follows.\begin{eqnarray}
BER & \approx & \frac{J}{n}P(J)+\frac{J}{n}{\displaystyle \sum_{i=J+1}^{N_{0}-1}|E_{J}|\binom{n-J}{i-J}}\varepsilon^{i}(1-\varepsilon)^{n-i}+\frac{M}{n}|E_{J}|\binom{n-J}{N_{0}-J}\epsilon^{N_{0}}(1-\epsilon)^{n-N_{0}}\nonumber \\
 & + & \frac{M}{n}\left[1-\sum_{i=0}^{N_{0}}\binom{n}{i}\varepsilon^{i}(1-\epsilon)^{n-i}\right].\label{eq:BER_opt}\end{eqnarray}

In the following we provide the rationale behind the derivation of
(\ref{eq:BER_opt}). We recall that $|E_{i}|\approx|S_{i}^{'}|$.
We partition $S_{i}^{'}$ further as $S_{i}^{'}=S_{i}^{''}\cup\overline{S_{i}^{''}}$,
where $S_{i}^{''}$ is the set of error patterns of weight $i$ that
have one and only one element of the set $E_{J}$ as their sub-pattern.
Using similar discussions as those used for the derivation of (\ref{eq:5}),
we have\begin{equation}
|S_{i}^{''}|\approx|S_{i}|\binom{i}{J}\left(1-\frac{|E_{J}|}{|S_{J}|}\right)^{\binom{i}{J}-1}\frac{|E_{J}|}{|S_{J}|}.\label{eq:P(S''_i))}\end{equation}

It appears that for the error patterns of weight $J\leq i<N_{0}$,
the ratio of $|S_{i}^{''}|/|\overline{S_{i}^{''}}|$ is a large number
and thus a large majority of the error patterns in the set $E_{i}$
belong to the set $S_{i}^{''}$. This means that, for this range of
error weights, many of the error patterns that result in decoding
failures have one and only one element of the set $E_{J}$ as their
sub-pattern. For these error patterns we make the following assumption:

\emph{Assumption (c)}: \emph{For the error patterns of weight $J\leq i<N_{0}$,
the number of bit errors at the end of decoding is approximately $J$.}

Furthermore, we assume:

\emph{Assumption (d): For the error patterns of weight $N_{0}\leq i\leq n$,
the number of bit errors at the end of decoding is on average $M$,
where $M$ is the estimate of the average number of bit errors for
error patterns of weight $N_{0}$.} 

To determine $M$, we simulate a given number, say $10^{5}-10^{6}$,
of randomly generated error patterns of weight $N_{0}$.

Our experiments show that both Assumptions (c) and (d) are statistically
viable and provide very good estimates of the total BER.

\subsection{Computational Complexity}

To obtain an estimate of FER, one needs to enumerate $\sum_{i=1}^{J}\binom{n}{i}$
error patterns. For different error patterns, the iterative decoder
performs the decoding and counts the number of decoding failures.
Suppose that we are interested in estimating the performance of the
LDPC code at a FER of $p$ using Monte Carlo simulations and would
like to observe at least $m$ codeword errors for a reliable result.
Assuming that the average number of computations required for iterative
decoding in the two cases are the same, the ratio of the computational
complexities of the Monte Carlo simulation and the proposed estimation
method is:

\begin{equation}
\eta=m/p{\displaystyle \sum_{i=1}^{J}\binom{n}{i}}.\label{eq:ComplexityFER}\end{equation}
For BER estimation, we need to perform an extra number of iterative
decodings to estimate $M$, and about $100m$ iterative decoding to
obtain $N_{0}$. These are usually negligible compared to $\sum_{i=1}^{J}\binom{n}{i}$. 

It can be seen that $\eta$ is a function of $n$, $J$, $p$ and
$m$. It increases with increasing $m$ and with decreasing $p$,
$n$ and $J$. It often appears that $J$ is a small number ($\leq4$).
For example, we have tested all the codes with rate $1/2$ and $n\leq2048$
in \cite{key-11} under GA, and they all have $J\leq4$. The value
of $m$ is often selected in the range of a few tens to a few hundreds.
With given values for $J$, $m$ and $n$, the proposed estimation
method is more efficient than the Monte Carlo simulation $(\eta>1)$
if $p<m/{\textstyle \sum_{i=1}^{J}\binom{n}{i}}$. In fact, for the
block length and $\varepsilon$ values of interest, the computational
complexity of the proposed method can be much smaller than that of
the Monte Carlo simulations.

One should also note that the value of $\eta$ in (\ref{eq:ComplexityFER})
only reflects the saving in complexity compared to \emph{one} Monte
Carlo simulation point. Unlike our estimations that can be easily
calculated for different values of $\varepsilon$ once we obtain $|E_{J}|$,
in Monte Carlo simulations, for each simulation point, a new set of
input vectors has to be generated and simulated. This makes the proposed
method even more attractive from the complexity viewpoint.

\section{Simulation Results}

To show that our method can be applied to both regular and irregular
LDPC codes and to a variety of hard-decision iterative decoding algorithms,
in this section, we perform experiments on four pairs of code/decoding
algorithm. Code 1 is a (200, 100) irregular LDPC code. The degree
distributions for this code, which are optimized for the BSC and GA,
are given by $\lambda(x)=0.1115x^{2}+0.8885x^{3}$ and $\rho(x)=0.26x^{6}+0.74x^{7}$
\cite{key-5}. Code 2 is a (210, 35) regular LDPC code which has a
variable node degree 5 and check node degree 6. For the degree distribution
of this code, MB algorithm of order zero (MB$^{0}$) has a better
threshold than GA does \cite{key-7}. Code 3 is a (1008, 504) regular
LDPC code with variable node degree 3 and check node degree 6 taken
from \cite{key-11}. Code 4, which is a (1998, 1776) regular LDPC
code taken from \cite{key-11}, has variable node degree 4 and check
node degree 36. Tanner graphs for all the codes are free of cycles
of length 4. Except Code 2, which is decoded by MB$^{0}$, all the
codes are decoded by GA. In our simulations, the maximum number of
iterations is 100 for all the decoders and for each crossover probability,
we simulate until we obtain 100 codeword errors.

Table \ref{Table:Categorization-of-Error} shows the values of $J$
and $|E_{J}|$ for the four codes. It also shows the number and the
percentages of different types of decoding failures for $E_{J}$.
As can be seen, for all the codes, most of the decoding failures caused
by initial error weight $J$ have fixed patterns and there is no random-like
failure.

\begin{figure}[h]
\begin{spacing}{0}
\begin{center}\includegraphics[%
  scale=0.75]{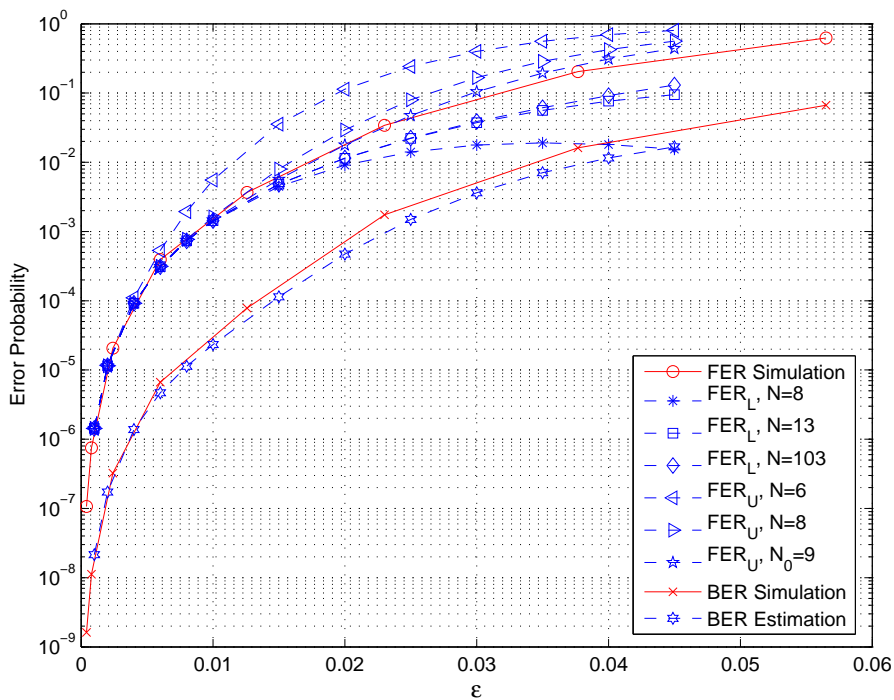}\end{center}
\end{spacing}

\begin{spacing}{0}

\caption{\label{cap:SimulationsCode2}FER and BER estimations and simulations
of the (200, 100) irregular code (Code 1, $\lambda(x)=0.1115x^{2}+0.8885x^{3}$,
$\rho(x)=0.26x^{6}+0.74x^{7}$) decoded by GA.}\end{spacing}

\end{figure}

\begin{figure}
\begin{spacing}{0}
\begin{center}\includegraphics[%
  scale=0.75]{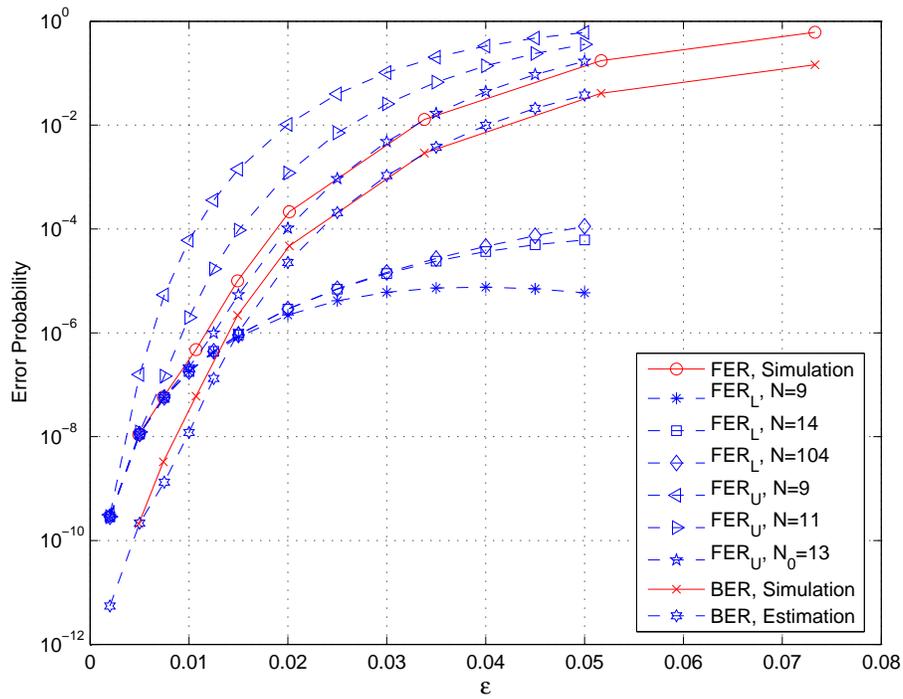}\end{center}
\end{spacing}

\begin{spacing}{0}

\caption{\label{cap:SimulationsCode3}FER and BER estimations and simulations
of the (210, 35) regular code (Code 2, $\lambda(x)=x^{4}$, $\rho(x)=x^{5}$)
decoded by MB$^{0}$. }\end{spacing}

\end{figure}

\begin{figure}[h]
\begin{spacing}{0}
\begin{center}\includegraphics[%
  scale=0.75]{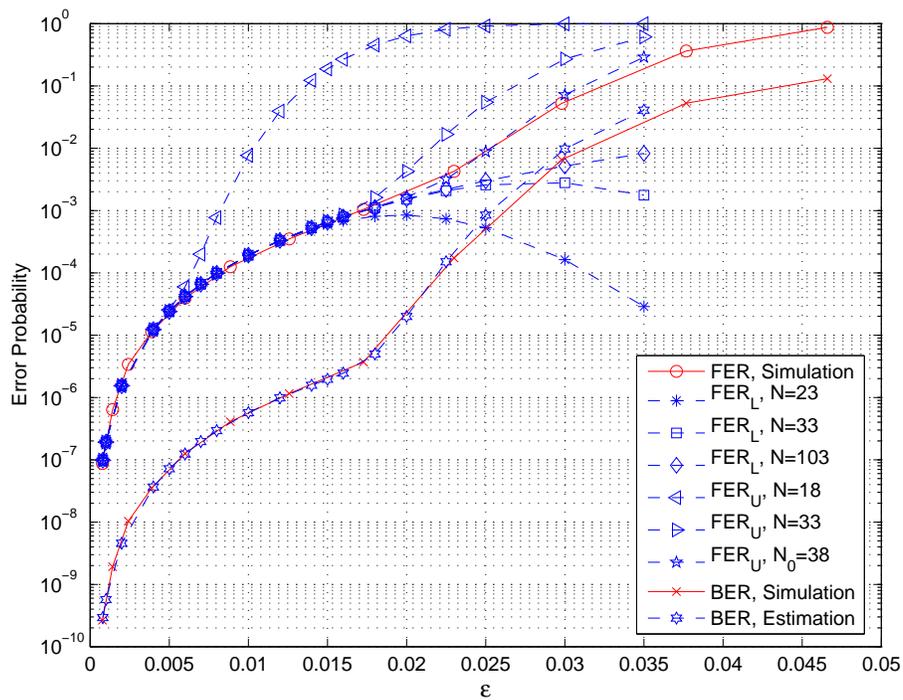}\end{center}
\end{spacing}

\begin{spacing}{0}

\caption{\label{cap:SimulationsCode4}FER and BER estimations and simulations
of the (1008, 504) regular code (Code 3, $\lambda(x)=x^{2}$, $\rho(x)=x^{5}$)
decoded by GA.}\end{spacing}

\end{figure}

\begin{figure}[h]
\begin{spacing}{0}
\begin{center}\includegraphics[%
  scale=0.75]{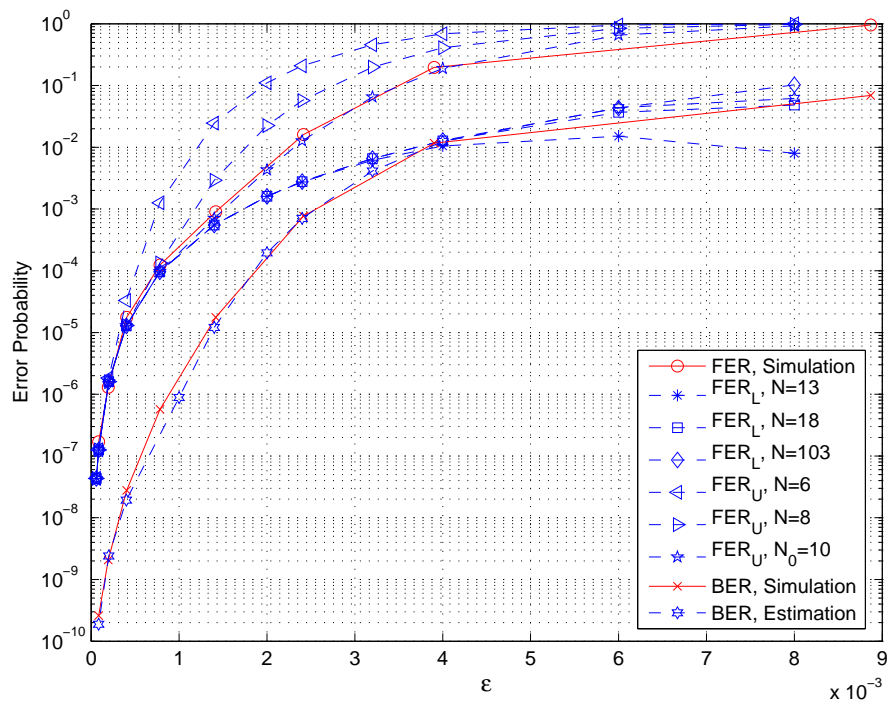}\end{center}
\end{spacing}

\begin{spacing}{0}

\caption{\label{cap:SimulationsCode5}FER and BER estimations and simulations
of the (1998, 1776) regular code (Code 4, $\lambda(x)=x^{3}$, $\rho(x)=x^{35}$)
decoded by GA.}\end{spacing}

\end{figure}

Figures \ref{cap:SimulationsCode2} - \ref{cap:SimulationsCode5}
show the FER and BER performances of Codes 1 - 4, obtained by simulations,
respectively. In these figures, we have also given the upper and the
lower estimates of the FER for different values of $N$, as well as
the estimates for the BER. For the estimates, we have used $N_{0}=9,13,38$
and $10$, for Codes 1 - 4, respectively. The corresponding values
of $M$ are $7.73,46.95,143.93$ and $133.57$, respectively. From
the figures, it can be seen that for each code, the lower estimate
$FER_{L}(N)$ improves by increasing $N$. However, even the best
estimate $FER_{L}(n)$ is only good at sufficiently small values of
crossover probability. It fails to provide an accurate estimate at
higher error rates. On the other hand, for all the codes, $FER_{U}(N_{0})$
provides an impressively accurate estimate of the FER over the whole
range of crossover probabilities of practical interest. It can also
be seen that the BER estimates for all the codes follow the simulated
BER curves very closely.

To compare the computational complexity of our proposed method with
that of the conventional Monte Carlo simulations, we consider the
following example.

\begin{example}
\label{exa:Complexity}We consider estimating the performance of Code
1 at $p=10^{-7}$. In this case, $J=3$, $n=200$ and we assume $m=100$
as is the case for the simulation results presented in Fig. \ref{cap:SimulationsCode2}.
With these values, we have $\eta=750$. If we were to add another
simulation point at $p=10^{-8}$, the complexity of Monte Carlo simulations
would increase to 8250 times that of our proposed method. For the
larger block length of 1008, our proposed method is more efficient
than Monte Carlo simulations if the target FER $p<6\times10^{-7}$.

One should note that, as we discussed in Section III-C, for any given
$n$ and $J$, there exists a FER $p'$, for which our method is more
efficient than the Monte Carlo simulations in estimating the FER over
the range $p<p'$. 
\end{example}

\section{Conclusions}

In this paper, we propose a method to estimate the error rate performance
of LDPC codes decoded by hard-decision iterative decoding algorithms
over a BSC. By only enumerating the smallest weight ($J$) error patterns
that cannot be all corrected by the decoder, the proposed method estimates
both FER and BER for a given LDPC code over the whole crossover probability
region of interest with very good accuracy. The proposed method is
universal in that it is applicable in principle to both regular and
irregular LDPC codes of arbitrary degree distributions and to any
hard-decision iterative algorithm. This universality is partly due
to the fact that, unlike previous approaches, our method is not based
on identifying the trapping sets and their relationship with the Tanner
graph structure of the code and the decoding algorithm.

Although the complexity of our proposed method is much less than that
of the Monte Carlo simulations for many cases of interest, it still
increases exponentially with $J$, essentially as $n^{J}$, where
$n$ is the block length. As a direction for future research, it would
be interesting to find (to estimate) the number of error patterns
of weight $J$ that cannot be corrected by the decoder, using methods
other than direct enumeration.

\begin{table}[h]

\caption{\label{Table:Categorization-of-Error}Categorization of {\scriptsize Error}
Patterns with Weight $J$ for Different LDPC Codes.}

\begin{singlespace}
\begin{center}\begin{tabular}{|c||c||c||c||c|}
\hline 
&
Code 1&
Code 2&
Code 3&
Code 4\tabularnewline
\hline
\hline 
$J$&
3&
4&
3&
3\tabularnewline
\hline 
$|E_{J}|$&
1434&
18&
192&
200479\tabularnewline
\hline 
Undetected Error&
0&
0&
0&
0\tabularnewline
\hline 
Fixed-Pattern&
1303(90.87\%)&
14(77.8\%)&
165(85.9\%)&
200438(99.98\%)\tabularnewline
\hline 
Oscillatory-Pattern&
131(9.13\%)&
4(22.2\%)&
27(14.1\%)&
41(0.02\%)\tabularnewline
\hline 
Random-like&
0&
0&
0&
0\tabularnewline
\hline
\end{tabular}\end{center}\end{singlespace}

\end{table}

\end{document}